\documentclass[prl,twocolumn,showpacs,floatfix,aps]{revtex4}
\usepackage{amssymb}
\usepackage{amsmath}
\usepackage{graphicx,bm}

\begin{document}

\title{\textbf{Hydrostatic-pressure effects on the pseudogap in slightly doped $YBa_2Cu_3O_{7-\delta}$ single crystals}}

\author{A.\,L.\,Solovjov$^{1,2}$, L.\,V.\,Omelchenko$^{1,2}$, R.\,V.\,Vovk$^3$, O.\,V.\,Dobrovolskiy$^{3,4}$ and
D.\,M.\, Sergeyev$^5$}
\email{Dobrovolskiy@Physik.uni-frankfurt.de}
\affiliation{$^1$B.\,I.\,Verkin Institute for Low Temperature Physics and Engineering of National Academy of Science of Ukraine, 47 Lenin ave., 61103 Kharkov, Ukraine\\
$^2$International Laboratory of High Magnetic
Fields and Low Temperatures, 95 Gajowicka Str., 53-421, Wroclaw, Poland\\
$^3$Physics Department, V. Karazin Kharkiv National University, Svobody Sq. 4, 61077 Kharkiv, Ukraine\\
$^4$Physikalisches Institut, Goethe University, Max-von-Laue-Str. 1, 60438 Frankfurt am Main, Germany\\
$^5$Zhubanov Aktobe Regional State University, Department of Condensed Matter Physics,
34A Moldagulova ave., 030000 Aktobe, Kazakhstan}


\begin{abstract}
The influence of hydrostatic pressure up to P=1.05~GPa on resistivity, excess conductivity $\sigma'(T)$
and pseudogap \, $\Delta^*(T)$ is investigated
in slightly doped single crystals of $YBa_2Cu_3O_{7-\delta}$ ($T_c(P=0) \approx$49.2~K\, and $\delta~\approx$~0.5).
The critical temperature $T_c$ is found to increase with increasing pressure at a rate $dT_c/dP = +5.1~KGPa^{-1}$,
while  $\rho(300)\,K$ decreases at a rate $dln\rho/dP = (- 19\pm 0.2)\%~GPa^{-1}$.
Near $T_c$, independently on pressure, the $\sigma'(T)$ is well described by the Aslamasov-Larkin and
Hikami-Larkin fluctuation theories, demonstrating a 3D-2D crossover with increase
of temperature.
The crossover temperature $T_0$ determines the coherence length along the c-axis $\xi_c(0)\simeq(3.43\pm0.01)$\AA\,
at P=0, which is found to decrease with increasing P.
At the same time, $\Delta^*$ and the BCS ratio $2\Delta^*/k_B\,T_c$ both increase with increasing hydrostatic pressure at a rate $dln\Delta^*/dP\approx 0.36\,GPa^{-1}$, implying an increase of the coupling strength with increasing P.
At low temperatures below $T_{pair}$, the shape of the $\Delta^*(T)$ curve is found to be almost independent on pressure.
At high temperatures, the shape of the $\Delta^*(T)$ curve changes noticeably with increasing P, suggesting a strong influence of pressure on the lattice dynamics.
This unusual behavior is observed for the first time.

\end{abstract}


\maketitle

$\bf{I.\, INTRODUCTION}$\\

The pseudogap (PG), which is opening in the excitation spectrum at the characteristic temperature $T^\ast \gg T_c$, remains to be one of the most interesting and intriguing property of high-temperature superconductors (HTSCs) with the active CuO$_2$ plane (cuprates) \cite{Kord,S1,PB}.
According to the definition proposed by Mott \cite{M,M2}, PG is a specific state of matter with a reduced density of the quasiparticle states (DOS) at the Fermi level at temperatures $T^\ast > T \gg T_c$,
where, $T_c$ is the superconducting transition temperature.
In YBa$_2$Cu$_3$O$_{7-\delta}$ (YBCO) a noticeable reduction of DOS at $T<T^\ast$, i.e. PG, was observed soon after the discovery of the cuprates by the measurement of the Knight shift, $K(T)$, i.\,e. the frequency shift in the nuclear magnetic resonance (NMR) \cite{Al}.
NMR measurements allow one to deduce the spin susceptibility $\chi_s(\omega,\mathbf{k})$ of the charge carriers.
The Knight shift being proportional to the spin polarization in external magnetic field determines the static ($\omega = 0$) and homogeneous ($\mathbf{k} = 0$) parts of the susceptibility, that is $K\sim \chi_s\equiv \chi_s(0,0)$.
In the Landau theory \cite{LP} $\chi_s(0,0) \backsim \rho_n(0)\equiv\rho_f$, where $\rho_n(\varepsilon)$ is the dependence of the density of the Fermi states on the energy in the normal phase.
In the classical superconductors $\rho_n(\varepsilon)$ (DOS) and, hence, $K(T)$ remains nearly constant in the whole temperature range of the existence of the normal phase, whereas in HTSCs it rapidly decreases at $T\leq T^\ast$ \cite{Al}.
Recently, the reduction of DOC and PG at $T<T^\ast = 170$\,K have been directly measured by angle resolved photoemission spectroscopy (ARPES) for the cuprate Bi2201 \cite{Kon}.
It was observed that as in the case of classical metals, DOS does not depend on temperature above $T^\ast$, but it starts to rapidly decrease at $T < T^\ast$.
In consequence of this, a depleted DOS, i.\,e. PG is observed in a broad temperature range from $T^\ast$ to $T_c = 32$\,K.
However, the physics of the processes leading to the decrease of DOS at $T^\ast > T \gg T_c$ remains uncertain so far \cite{Kord,S1,PB,Tel}.

There is a noticeable number of theoretical models addressing the non-superconducting nature of the appearance of PG, see e.\,g. Refs. \cite{Kord,Tel,Gab,Nor,Ber} and references therein.
However, we adhere another viewpoint that PG appears in consequence of the formation of paired fermions (local pairs) in HTSCs at $T\leq T^\ast$,  see e.\,g. Refs. \cite{S1,EK,Cho,Yaz,Mis,Tch,Kag} and references therein.
According to the theories of systems with small charge carrier density $n_f$ \cite{L,H,Eng}, right those as HTSCs are, the local pairs (LP) can appear at $T\leq T^\ast$ in the form of the so-called strongly-bound bosons (SBB).
By definition, SBB are low-dimensional, but exceptionally strongly bound pairs obeying the theory of Bose-Einstein condensation (BEC).
The pair size is determined by the coherence length in the $ab$-plane, $\xi_{ab}$, the typical value of which in YBCO with a close-to-optimal doping level $\xi_{ab} \sim (5 - 10)$\AA\, \cite{Sug,WzK}.
Accordingly, the bound energy in this pair, $\varepsilon_b \sim 1/(\xi_{ab})^2$, is very large \cite{H,Eng}.
In consequence of this, SBB are not destroyed by thermal fluctuations and do not interact one with another since the pair size is much smaller than the distance between them.
However, LP can only condensate at $T_c \ll T^\ast$ \cite{L,H,Eng}.
For this reason, upon approaching $T_c$ SBB have to transform into fluctuating Cooper pairs (FCP) which obey the Bardeen-Cooper-Schrieffer (BCS) theory \cite{DeGen}.
In this way, the theory predicts the BEC-BCS transition with decreasing $T$, as observed experimentally \cite{Kon2,ST}.
In YBCO thin films, the temperature of this transition amounts to $T_{pair}\sim 130$\,K \cite{S2}.
At the same time, there exists one more characteristic temperature $T_c < T_{01} < T_{pair}$.
According to the theory, the wave function phase stiffness has to be maintained up to $T_{01}$ \cite{EK,Cho}.
This means that the superfluid density, $n_s$, maintains a nonzero value up to $T_{01}$ \cite{EK,Cor,Tal,DTS}.
However, the details of the BEC-BCS transition are not fully clear so far as well \cite{S1,H,Eng,ST,Gus}.

Pressure is a powerful tool for studying various properties of the cuprates \cite{Liu,Wan,Fer,She,V1,V2,V3} and it is widely used in experiments since the discovery of HTSCs \cite{Chu} till present days \cite{Fan}.
Pressure noticeably affects $T_c$ and the resistance of HTSCs in the normal state.
In contrast to the conventional superconductors, in the cuprates in the vast majority of cases the dependence $dT_c/dP$ is positive, whereas the derivative $d\ln\rho_{ab}/dT$ is negative and relatively large \cite{Liu,Wan,Fer}.
Here, $\rho_{ab}$ is the resistivity in the $ab$-plane, that is parallel to the CuO$_2$ conducting layers.
The pressure impact mechanisms on $\rho$ are not ultimately understood for the reason that the nature of the transport properties of HTSCs, strictly speaking, is not completely clear.
As is well known, the main contribution to the conductivity of the cuprates is provided by the $CuO_2$ planes between which there is a relatively weak interlayer interaction.
Pressure is likely to lead to a redistribution of the charge carriers and to an increase of their concentration $n_f$  in the conducting CuO$_2$ planes that should lead to a reduction of $\rho$.
Properly, the increase of $n_f$ under pressure should also lead to an increase of $T_c$, i.e. to a positive value of $dT_c/dP$ observed in experiment.
This process should take place easier in slightly doped samples \cite{S4}, where $n_f$ is small and there is a large number of oxygen vacancies \cite{Cav,Ast}.

The theoretical problem of the effect of hydrostatic pressure on $\rho_{ab}$ in HTSCs was addressed in Ref. \cite{Liu}.
There are also several works where the influence of pressure on the fluctuation conductivity (FLC) in various cuprates was studied \cite{Wan,Fer,She}.
Recently, it has been shown that pressure also noticeably increases the value of the superconducting gap in various cuprates \cite{Kha,DT}.
At the same time, there has been few works addressing the pressure effect on PG \cite{S4,S5}.

Here, we investigate of the hydrostatic pressure effect on the temperature dependencies of the resistivity $\rho_{ab}(T)$ in slightly doped $YBa_2$Cu$_3$O$_{7-\delta}$ single crystals (YBCO) with $T_c = 49.2$\,K at $P=0\,GPa$.
We investigated the fluctuation contributions to the conductivity, chiefly focusing on the temperature dependence of the excess conductivity $\sigma^\prime(T)$.
\begin{figure}[t]
\begin{center}
\includegraphics[width=.48\textwidth]{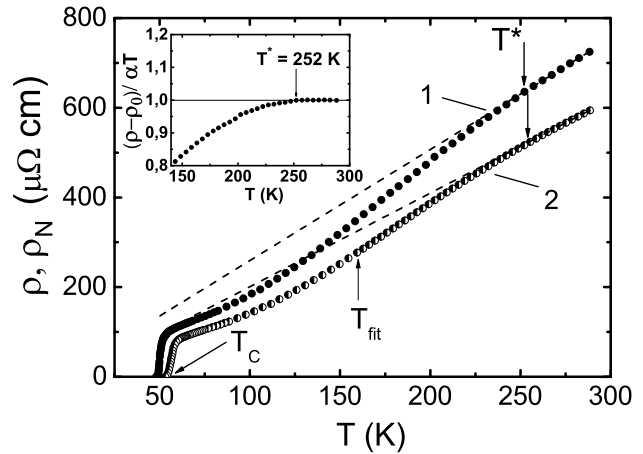}
\caption{Temperature dependence of $\rho$ of $YBa_2Cu_3O_{7-\delta}$ ($7-\delta\simeq6.5$) single crystal at P=0\,GPa (curve 1, dots,) and 1.05~GPa (curve 2, semicircles). Dashed lines depict the extrapolations of $\rho_N(T)$ to the low T region. $T_{fl}$ is a temperature down to which the polynomial fitting was performed.
Inserts display the determination of $T^*$ at P=0\,GPa using the $(\rho(T)-\rho_0))/\alpha T$ criterion (see text for details).}
\end{center}
\end{figure}
From the analysis of the excess conductivity the value and the temperature dependence of FLC
and the pseudogap $\Delta^\ast(T)$ at pressures up to $Ð=1.05$\,GPa ($1$\,GPa=$10$\,kbar) are obtained.
The analysis is conducted in the framework of our model of the local pairs \cite{S1,ST}, as detailed in the text.
Comparison of our results with the results obtained for Bi$_2$Sr$_2$CaCu$_2$O$_{8-\delta}$ (BiSCCO-2212) \cite{Liu}, $HgBa_2Ca_2Cu_3O_8$ (Hg-2223) \cite{She} and slightly doped  HoBa$_2$Cu$_3$O$_{7-\delta}$ \cite{S4} should help to understand better the mechanisms of the pressure effect on $T_c$, $\rho_{ab}(T)$, FLC and $\Delta^\ast(T)$.\\

\indent {\bf II.\, EXPERIMENT}\\

The YBa$_2$Cu$_3$O$_{7-\delta}$ (YBCO) single crystals were grown by the solution-melt technique according to Refs. \cite{V1,V2,V3,V4}.
For electrical resistance measurements were selected crystals of rectangular shape with typical dimensions of $3\times5\times0.3$\,mm$^3$.
The minimal dimension corresponds to the $c$-axis.
To obtain sample with a given oxygen content, the crystals were annealed in an oxygen atmosphere as described in Refs. \cite{V1,V4}.
The electrical resistance in the $ab$-plane was measured in the standard four-probe geometry with a dc current up to 10\,mA \cite{V5} in the regime of fully automated data acquisition.
The measurements were conducted in the temperature sweep mode, with a rate of  $0.1$\,K/min near $T_c$ and about $5$\,K/min at $T\gg T_c$.

Hydrostatic pressure was created in an autonomous chamber of the cylinder-piston type according to the technique described in Refs. \cite{V5,Tho}.
For the determination of the effect of the oxygen redistribution the measurements were conducted in two to seven days after the application of pressure, after the relaxation processes had been completed \cite {S4}.
Fig. 1 displays the temperature  dependencies of the resistivity $\rho(T) \equiv \rho_{ab}(T)$ of the $YBa_2$Cu$_3$O$_{7-\delta}$ single crystal with $T_c(P=0) = 49.2$\,K and the oxygen index $7-\delta\sim6.5$ \cite{Ito} measured at $P=0\,GPa$ (curve 1) and $P = 1.05$\,GPa (curve 2).
The curves have an expected $S$-shaped form typical for slightly doped YBCO films \cite{S1,Mo} and single crystals \cite{An,Ito}.
\begin{figure}[t]
\begin{center}
\includegraphics[width=.54\textwidth]{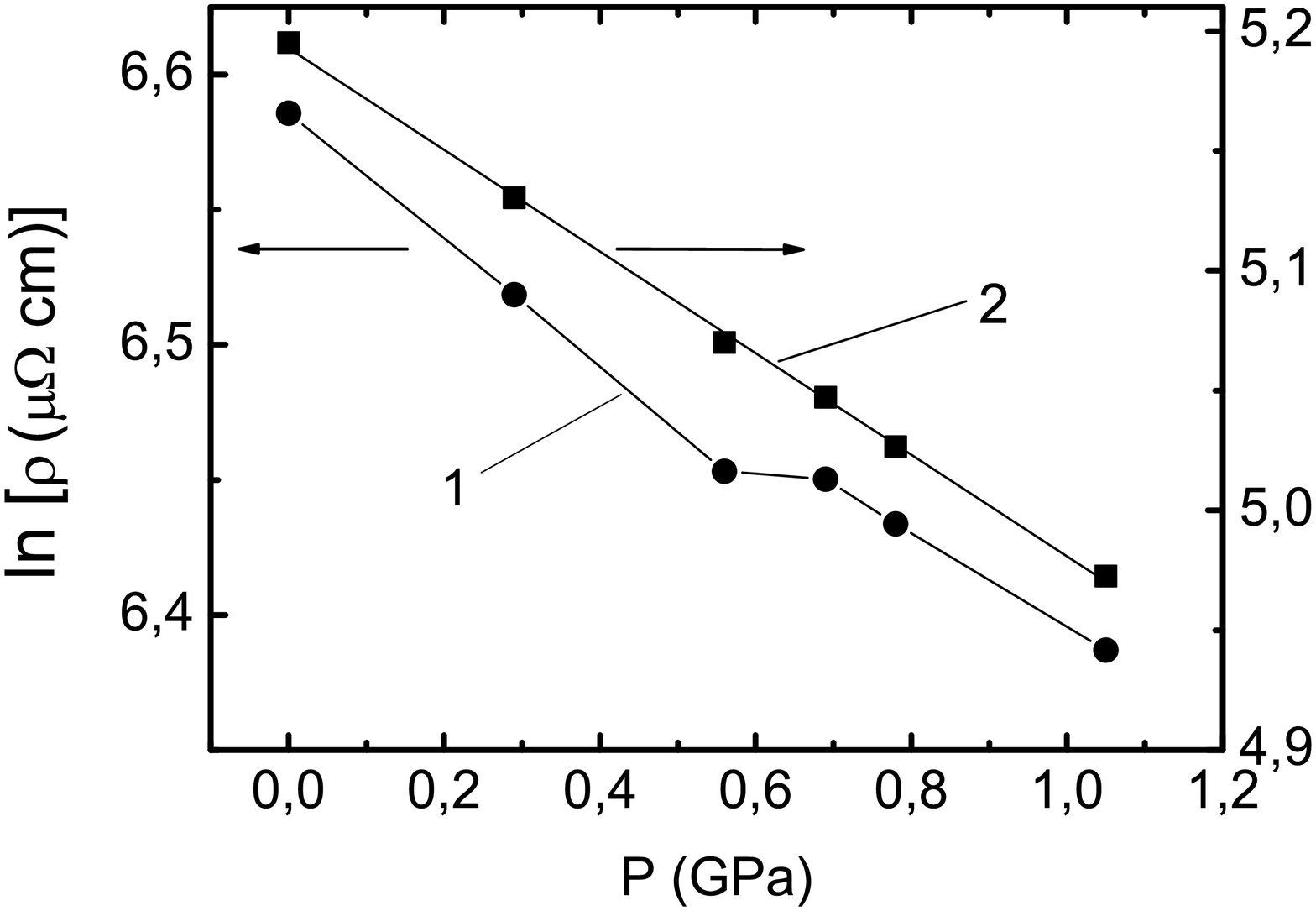}
\caption{Pressure dependence of $ln\rho$ of the
$YBa_2Cu_3O_{6.5}$ single crystal at 288 K (curve 1, dots). Solid line is guide for the eye.
Squares (curve 2) represent $ln\rho(P)$ measured at T=100\,K along with the
least-squares fit.}
\end{center}
\end{figure}

Besides, as it follows from the theory \cite{Tal,Tch,L,H,Eng}, in this case the values of the characteristic temperature $T^\ast$ are noticeably higher than $T^\ast\sim 140$\,K observed in optimally doped YBCO compounds \cite{Ito,Mo}.
Nevertheless, in the temperature range from $T^\ast = (252\pm0.5)$\,K (curve 1) and $T^\ast = (254\pm0.5)$\,K (curve 2) to $\simeq 300$\,K, the dependence $\rho(T)$ is linear with the slopes $d\rho/dT = 2.48\,\mu\Omega$cm/K$^{-1}$ and $d\rho/dT = 2.08\,\mu\Omega$cm/K$^{-1}$ for $P=0\,GPa$ and $P= 1.05$\,GPa, respectively (Fig. 1).
The slopes were determined by computer linear fitting which confirms a rather good linearity of the dependences in the stated temperature range with the standard error of about $0.009\pm 0.002$ at all applied pressures.
The PG temperature $T^*$ is taken at the point where the experimental resistivity
curve starts to turn down from the linear high-temperature behavior depicted by the dashed lines in the figure.

A more precise approach for the determination of $T^*$ relies upon the criterion $[\rho(T)-\rho_0]/aT$ \cite{DeM}.
Now $T^*$ is the temperature at which $[\rho(T)-\rho_0]/aT$ turns down from 1 as shown in the insert in Fig. 1.
Both approaches yield the same T*'s values.
In fact, we have six curves (six samples:Y0 - Y6) measured at P=0, 0.29, 0.56, 0.69, 0.78 and 1.05 GPa.
The sample parameters obtained at different P are listed in Tables I and II.
The dependencies $\rho(T)$ measured for all intermediate pressure values also have the $S$-shaped form and are located between the two curves shown in Fig. 1.
The whole set of curves resembles that shown in Fig. 1 in Ref. \cite{S4}.

As can be seen from Table II, the pressure actually does not affect the $T^*$ values .
At the same time, the linear slope $a$ is found to linearly decrease with P at a rate $da/dP=(0.38\pm0.02)$ $\mu \Omega cm K^{-1}GPa^{-1}$.
Simultaneously the pressure increase leads to a noticeable reduction of the resistance of the sample.
The relative reduction of $\rho(T)$ as a function of pressure is practically independent on temperature above $260$\,K
and amounts to $d\ln\rho(300\,K)/dP =(- 19\pm0.2)\%\,GPa^{-1}$ (Fig 2, curve 1).
This value is smaller than $d\ln\rho/dP = (- 25.5 \pm 0.2)\%$\,GPa$^{-1}$ for BiSCCO single crystals \cite{Liu}, but it is noticeably larger than $d\ln\rho/dP = (- 4 \pm 0.2)\%$\,GPa$^{-1}$ obtained by us for slightly doped HoBCO single crystals \cite{S4}.

At the same time $d\ln\rho(100\,K)/dP$ amounts to $(- 14.8\pm0.2)\%$\,GPa$^{-1}$ (Fig 2, curve 2).
The typical value for YBCO single crystals $d\ln\rho/dP = (- 12 \pm 0.2)\%$\,GPa$^{-1}$ is in good agreement with our results, given the different doping level of the samples (see Ref. \cite{Liu} and references therein).
We note that $d\ln\rho(100\,K)/dP$ demonstrates a nearly linear dependence on P with a standard error of about 0.00323 (Fig 2, curve 2) which is typical for the monocrystalline cuprates \cite{Liu}.
In contrast, $d\ln\rho(300\,K)/dP$ of the studied YBCO single crystal displays the noticeable deviation from linearity centered at $\sim 0.7\,GPa$ (Fig. 2, curve 1).
The peculiarity is also seen in the PG results, as will be discussed in a follow-up paragraph.
\begin{figure}[b]
\noindent\centering{
\includegraphics[width=.68\textwidth]{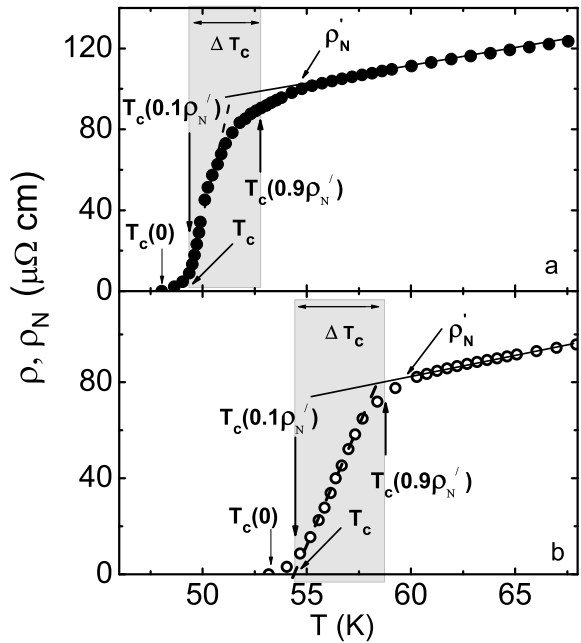}}
\caption{Resistive transitions of $YBa_2Cu_3O_{6.5}$ single crystal at P=0\,GPa (panel a, dots) and P=1.05~GPa (panel b, circles).
$T_c$ is determined by extrapolation of the linear part of the resistive transition to $\rho(T_c)=0$ (dashed lines in the figure).
Solid lines designate the linear $\rho(T)$ regions just above the onset temperature $T_{on}$, which
defines $\rho_{N}'$. The shaded areas determine the width of the resistive transitions
$\Delta T_c = T_c(0.9\rho_{N}') - T_c(0.1\rho_{N}')$.}
\label{ResTr}
\end{figure}
In spite of a number of studies of the relaxation processes in the 1-2-3 system under high pressure, many aspects, such as the charge transfer and the nature of redistribution of the vacancy subsystem, still remain uncertain \cite{Liu,Fer,V1,V2,V3,S4,V4}.
Thus, the electrical resistivity decreases not only as a consequence of the high pressure, but also in the isobar process of retaining the sample at room temperature, following the application of pressure \cite{S4,V2,V3}.
Importantly, when the pressure is removed, $\rho(T)$ finally coincides with the original curve obtained before the application of pressure \cite{S4}.
This experimental fact confirms the reversibility of the process.

Figure 3 displays the resistive curves in the vicinity of $T_c$ for $P=0\,GPa$ (a) and $P= 1.05$\,GPa (b), respectively, which contain all characteristic temperatures of the superconducting (SC) transition.
As usually, the transition temperature $T_c$ is determined by extrapolation of the linear part of the resistive transition (dashed lines in Fig. 3) to $\rho(T_c)=0$ \cite{Lang} .
It is seen that the resistive transitions are rather broad: $\Delta T_c = T_c(0.9\rho'_N) - T_c (0.1\rho'_N) = (52.77 - 49.45)\,\mathrm{K}=3.32\,\mathrm{K}\,(P=0\,GPa)$ and $\Delta T_c = 58.76-54.74=4$\,K ($P=1.05$\,GPa).
Here $T_c(0.9\rho'_N)$ and $T_c(0.1\rho'_N)$ correspond to the temperatures at which resistivity decreases 10\% and 90\%, respectively, with respect to the $\rho'_N$  value just above the resistive transition designated by the upper straight line (Fig. 3).
In this way, pressure broadens the resistive transition by about $20\%$, that is not so pronounced as in HoBCO single crystals where the effect is of the factor of $\simeq 2$ \cite{S4}.
In addition to this, one sees that $T_c$ expectedly rises from $49.2\,K$ up to $54.6\,K$ with increasing pressure, (see also Fig. 4).

From Fig. 3 and 4 we deduce that $T_c$ increases with increasing hydrostatic pressure at a rate $dT_c/dP\simeq +5.1$\,KGPa$^{-1}$ which is in a good agreement with our results for slightly doped (SD) HoBCO single crystals where $dT_c/dP\simeq  +4$\,KGPa$^{-1}$ \cite{S4}.
The same value $dT_c/dP\simeq +4$\,KGPa$^{-1}$ was also observed by pressure experiments in SD polycrystalline $YBa_2Cu_3O_{7-\delta}$ $(7-\delta\sim 6.6)$ by muon spin rotation ($\mu\,SR$) \cite{Mai}.
This result confirms the expressed assumption that in cuprates, $T_c$ is likely to rise at the expense of the increase of the charge carrier density $n_f$ in the $CuO_2$ planes under pressure.
Meanwhile, it is likely that the oxygen vacancies in slightly doped cuprates provide the possibility for a more easy redistribution of $n_f$ as compared with optimally doped samples where the number of vacancies is small and $n_f$ is, in  turn, rather large.\\

\indent {\bf III.\, RESULTS AND DISCUSSION}

\indent {\bf A.\, Fluctuation conductivity}\\

Independently on the value of the applied pressure,
below the PG temperature $T^*$
resistivity curves of studied $YBa_2Cu_3O_{6.5}$ single crystal turn down from the linear behavior of $\rho$(T) observed at higher temperatures (Fig.1).
This leads to appearance of the excess conductivity
\begin{equation}
\sigma '(T) = \sigma(T) - \sigma_N(T)=[1/\rho(T)]-[1/\rho_N(T)],
\label{sigma-t}
\end{equation}
where $\rho_N(T)$ = aT+$\rho_0$ is the linear normal state resistivity extrapolated to the low-$T$ region \cite{S1,SP} and $\rho_0$ is the intercept with the y-axis.
This procedure of the normal state resistivity determination is widely used in literature (see \cite{S1,DeM,Lang,Oh,ND} and references therein) and has been justified theoretically within the frame-work of the nearly antiferromagnetic Fermi liquid (NAFL) model \cite{SP}.
\begin{figure}[t]
\noindent\centering{
\includegraphics[width=.66\textwidth]{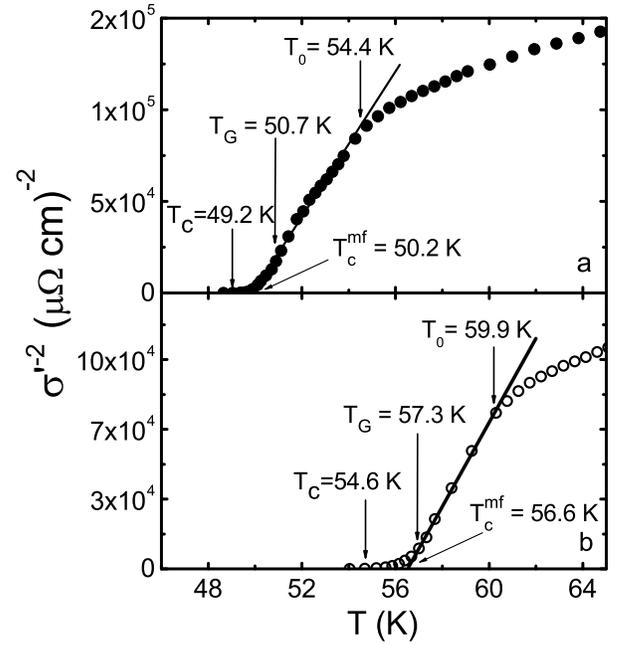}}
\caption{$\sigma'^{-2}$ as a function of T at P=0\,GPa (panel a, dots) and P=1.05~GPa (panel b, circles).
The straight lines follow the low $\sigma'^{-2}(T)$ which actually corresponds to the
3D AL fluctuation region. All temperatures designated by the arrows in the figure are discussed in the text.}
\label{sigma-e}
\end{figure}

Here we focus on the analysis of FLC and PG derived from measured excess conductivity within our LP model.
We mainly perform the analysis for the sample Y0 (P=0) and compare the results with those obtained for sample
Y6 (with P=1.05 GPa applied for five days) as well as with the results
obtained for BiSCCO, YBCO \cite{Liu,Kha,Mai} and HoBCO \cite{S4} single crystals.
Naturally, the same analysis has been performed for all other samples under study.
The sample parameters derived from the analysis at different values of pressure
are listed in Tables I and II.

\indent First of all, the mean field critical temperature $T_c^{mf}$ has to be found.
Here $T_c^{mf}>T_c$ is the critical temperature in the mean-field approximation, which separates the FLC region from the region of critical fluctuations or fluctuations of the SC order parameter $\Delta_0$ directly near $T_c$ (where $\Delta_0<kT)$, neglected in the Ginzburg-Landau (GL) theory \cite{LP,GL1}.
In all equations used in the analysis the reduced temperature \cite{HL}is used, viz.,
\begin{equation}
\varepsilon = (T - T_c^{mf})\,/\,T_c^{mf}.
\label{var}
\end{equation}
The correct determination of $T_c^{mf}$ is hence crucial for the FLC and PG calculations.


Within the LP model it was convincingly shown that FLC measured for all HTSCs always demonstrates a crossover from the 2D ($\xi_c(T)<d$) in 3D ($\xi_c(T)>d$) regime as T approaches $T_c$ (\cite{S1,Beas,Xie} and references therein).
As a result, near $T_c$ FLC is always extrapolated by the standard equation of the Aslamasov-Larkin (AL) theory \cite{AL} with the critical exponent $\lambda=-1/2$ (Fig. 5, dashed line 1) which determines FLC in any 3D system
\begin{equation}
\sigma_{AL3D} '=C_{3D}\frac{e^2}{32\,\hbar\,\xi_c(0)}{\varepsilon^{-1\,/\,2}},
\label{3D}
\end{equation}
where $\xi_c(T)$ is a coherence length along the c-axis, d is a distance between the conducting layers \cite{HL},
and $C_{3D}$ is a numerical factor to fit the data to the theory \cite{S1,Oh,Beas}.
This means that the conventional 3D FLC is realized in HTSCs as T approaches $T_c$ \cite{S1,Xie}.
The result is most likely a consequence of Gaussian fluctuations of the order parameter in 2D metals which were found to prevent any phase coherence organization in 2D compounds \cite{L,H,Eng}.
As a result, the critical temperature of an ideal 2D metal is found to be zero (Mermin-Wagner-Hoenberg theorem) and a finite value is obtained only when three-dimensional effects are taken into account \cite{Tch,L,H,Eng}.
From Eq. (3), one can easily obtain $\sigma'^{-2}\sim (T - T_c^{mf})\,/\,T_c^{mf}$.
Evidently,  $\sigma'^{-2}=0$ when $T = T_c^{mf}$ (Fig. 4).
This way of the determination of $T_c^{mf}$ was proposed by Beasley \cite{Beas} and justified in different FLC experiments \cite{S1,S5,Oh,ND}.
Moreover, when $T_c^{mf}$ is properly chosen, the data in the 3D fluctuation region near $T_c$ can always be fitted to Eq. (3).

\indent Fig. 4 a displays the $\sigma'^{-2}$ vs T plot (dots) for sample Y0 (P=0\,GPa).
The interception of the extrapolated linear $\sigma'^{-2}$ with the T-axis determines $T_c^{mf}= 50.2 ~K$ (Table I).
Now, when $T_c^{mf}$ is found, Eq.(2) allows one to determine $\varepsilon(T)$.
Above the crossover temperature $T_0 = 54.4~K$ ($ln\varepsilon_0=-2.45$, Fig. 5) the data deviate on the right from the line, suggesting the 2D Maki-Thompson (MT) \cite{Mak,Th} fluctuation contribution to FLC \cite{S1,HL,S6}.
Evidently, at the crossover temperature $T_0\sim \varepsilon_0$ the coherence length $\xi_c(T)=\xi_c(0)\varepsilon^{-1/2}$ is expected to amount to d \cite{ST,HL,S6}, which yields
\begin{equation}
\xi_c(0) = d\,\sqrt\varepsilon_0
\end{equation}
and allows one to determine $\xi_c(0)$ which
is one of the important parameters of the PG analysis.
Fig. 4b demonstrates the same consideration (circles) for sample Y6 (P=1.05\,GPa) which yields $T_c^{mf}= 56.6 ~K$.
Also shown in the figure is the representative temperature $T_G$.
It is this temperature which is generally accounted for by the Ginzburg criterion which is
related to the breakdown of the mean-field GL theory to describe the superconducting transition as mentioned above \cite{DeGen,GL1,Kap}.
Above $T_c$ this criterion is identified down to the lowest temperature
limit for the validity of the Gaussian fluctuation region.
In Fig. 4 and 5 we denote as $T_G$ the crossover temperature delimiting
the 3D Al fluctuation and critical intervals, and assign this temperature
to the point where the data deviate from the straight lines corresponding to the 3D AL regime.

\indent When $\varepsilon(T)$ is determined, the role of the fluctuating pairing in the PG formation can be clarified \cite{S1,Tch,L,H,Eng,HL}.
To accomplish this,  $ln\sigma'$ vs $ln\varepsilon$ is plotted in Fig. 5 (a --- P=0\,GPa, and b --- P=1.05\,GPa) in comparison with the fluctuation theories.
As expected, above $T_c^{mf}$ and up to $T_0$ = 54.5~K ($\ln \varepsilon_0 \approx -2.45$) $\sigma'$ vs $T$ is well extrapolated by the 3D fluctuation term by Eq. (3) of the AL theory (Fig. 5a, dashed line 1) with $\xi_c(0)= (3.43 \pm 0.02)$\AA\ determined by Eq. (4) and $C_{3D}\sim\,4.0$ (see Table I).
Besides, by analogy with YBCO films \cite{Mo,Lang,Oh,S6} and single crystals \cite{Liu,Ito,An}, we use d=11.67\AA=c which is the c-axis lattice parameter \cite{Chr}.
Accordingly, above $T_0$ and up to $T_{01}\approx  87.4$~K ($\ln \varepsilon_{01} \approx -0.3$) $\sigma'$ can be described well by the MT fluctuation term (5) (Fig. 5a, solid curve 2) of the Hikami-Larkin (HL) theory \cite{HL}
\begin{equation}
\sigma_{MT} '=\frac{e^2}{8\,d\,\hbar}\frac{1}{1-\alpha/\delta}\,ln\left((\delta/\alpha)\,
\frac{1+\alpha+\sqrt{1+2\,\alpha}}{1+\delta+\sqrt{1+2\,\delta}}\right)\,\varepsilon^{-1},
\label{MT}
\end{equation}
which dominates well above $T_c$ in the 2D fluctuation region \cite{S6,Xie,HL}.
In Eq. (5)
\begin{equation}
\alpha = 2\biggl[\frac{\xi_c(0)}{d}\biggr]^2\,\varepsilon^{-1}
\label{alp}
\end{equation}
is a coupling parameter,

\begin{equation}
\delta=\beta
\frac{16}{\pi\,\hbar}\biggl[\frac{\xi_c(0)}{d}\biggr]^2\,k_B\,T\,\tau_{\phi}
\label{TM}
\end{equation}
is the pair-breaking parameter,
and $\tau_{\phi}$ defined by the relation
\begin{equation}
\tau_{\phi}\beta\,T={\pi\hbar}/{8k_B\varepsilon}=A/\varepsilon,
\label{tau}
\end{equation}
is the phase relaxation time, and $A=2.998\cdot 10^{-12}$ sK.
The factor $\beta=1.203(l\,/\,\xi_{ab}$), where $l$ is the mean-free path and $\xi_{ab}$ is the coherence length in the {\it ab} plane in the clean limit ($l>\xi$) \cite{S1,S6}.
\begin{figure}[t]
\begin{center}
\includegraphics[width=.68\textwidth]{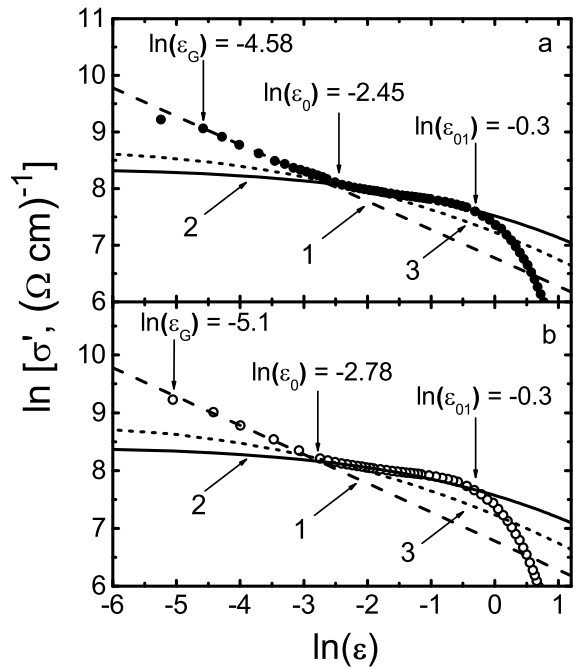}
\caption{$ln\sigma'$ vs $ln\varepsilon$ at P=0\,GPa (panel a, dots) and P=1.05~GPa (panel b, circles) compared with the fluctuation theories: 3D AL (dashed line 1); MT with $d=d_{1}$ (solid curve 2), and MT with $d=11.67$ \AA\, (short dashed curve 3).
$ln\varepsilon_{01}$ corresponds to $T_{01}$ which determines the range of the SC fluctuations, $ln\varepsilon_{0}$ corresponds to the crossover temperature $T_{0}$, and $ln\varepsilon_G$ designates the Ginzburg temperature $T_G$.}
\end{center}
\end{figure}

\indent Unfortunately, neither $l$ nor $\xi_{ab}(T)$ are accessible in our experiments.
To proceed with the analysis we will use the experimental fact that $\delta\approx 2$ when all other parameters are properly chosen \cite{S6}.
Thus, to calculate the MT fluctuation contribution using Eq. (5), only the coupling parameter $\alpha$ by Eq. (6) remains to be defined.
To determine $\alpha$ we have to use another experimental fact that
$\xi_c(0)= d \varepsilon_0^{1/2} = d_1 \varepsilon_{01}^{1/2} =
(3.43 \pm 0.02)$\AA\, \cite{S2,S4}.
Here $d_1$  corresponds to $T_{01}$ and is a distance between conducting $CuO_2$ planes in YBCO compounds.
Substituting d=11.67\,\AA \, one can easy obtain $d_1= d\sqrt{\varepsilon_0/\varepsilon_{01}}= 3.98\pm 0.05\,\AA\,$ which is actually the inter-planar distance in SD $YB_2Cu_3O_{6.65}$ at P=0\,GPa \cite{Chr}).
This finding suggests that $\varepsilon_{01}$ is properly chosen.
The same considerations performed for Y6 (P=1.05 GPA) provide a very similar $ln\sigma'$ vs $ln\varepsilon$ (Fig. 5b) with $\xi_c(0)= (2.91 \pm 0.02)$\AA\ and $d_1= (3.37 \pm 0.02)$\AA\ (Table I).

Both $ln\varepsilon_{01}$ and $ln\varepsilon_{0}$ as well as $ln\varepsilon_G$ are marked by the arrows in Fig. 5.
Within the LP model it is believed that below $\varepsilon_{01}$, the $\xi_c(T)$ exceeds $d_1$ and couples the $CuO_2$ planes by the Josephson interaction
resulting in the appearance of 2D FLC of the MT type which lasts down to $T_0$ \cite{S1,Xie}.
Thus, it turns out that only $\varepsilon_{01}$ has to govern Eq. (5) now and its proper choice is decisive for the FLC analysis.
As mentioned above, the corresponding temperature $T_{01}$ is introduced to determine the temperature range in which the SC order parameter wave function stiffness has to be maintained \cite{EK,Cho}.
As it is clearly seen from our analysis, it is just the range of the SC fluctuations which obey the conventional fluctuation theories.
That is why we have to substitute $\varepsilon_{01}$ instead of $\varepsilon_{0}$ into Eq. (8) to find $\tau_{\phi}$\,(100K)\,$\beta$ = (0.404\,$\pm$\,0.002)$\cdot$10$^{-13}$s.
If we use d=11.67\AA\ and set $\varepsilon$=$\varepsilon_{0}$ in Eq.(8), it results in $\tau_{\phi}$\,(100K)\,$\beta$= (3.47\,$\pm$\,0.002)$\cdot$10$^{-13}$s, and we get the curve 3 (Fig. 5) which does not meet the experimental case.
The result has to support the above considerations.
\begin{figure}[b]
\begin{center}
\includegraphics[width=.54\textwidth]{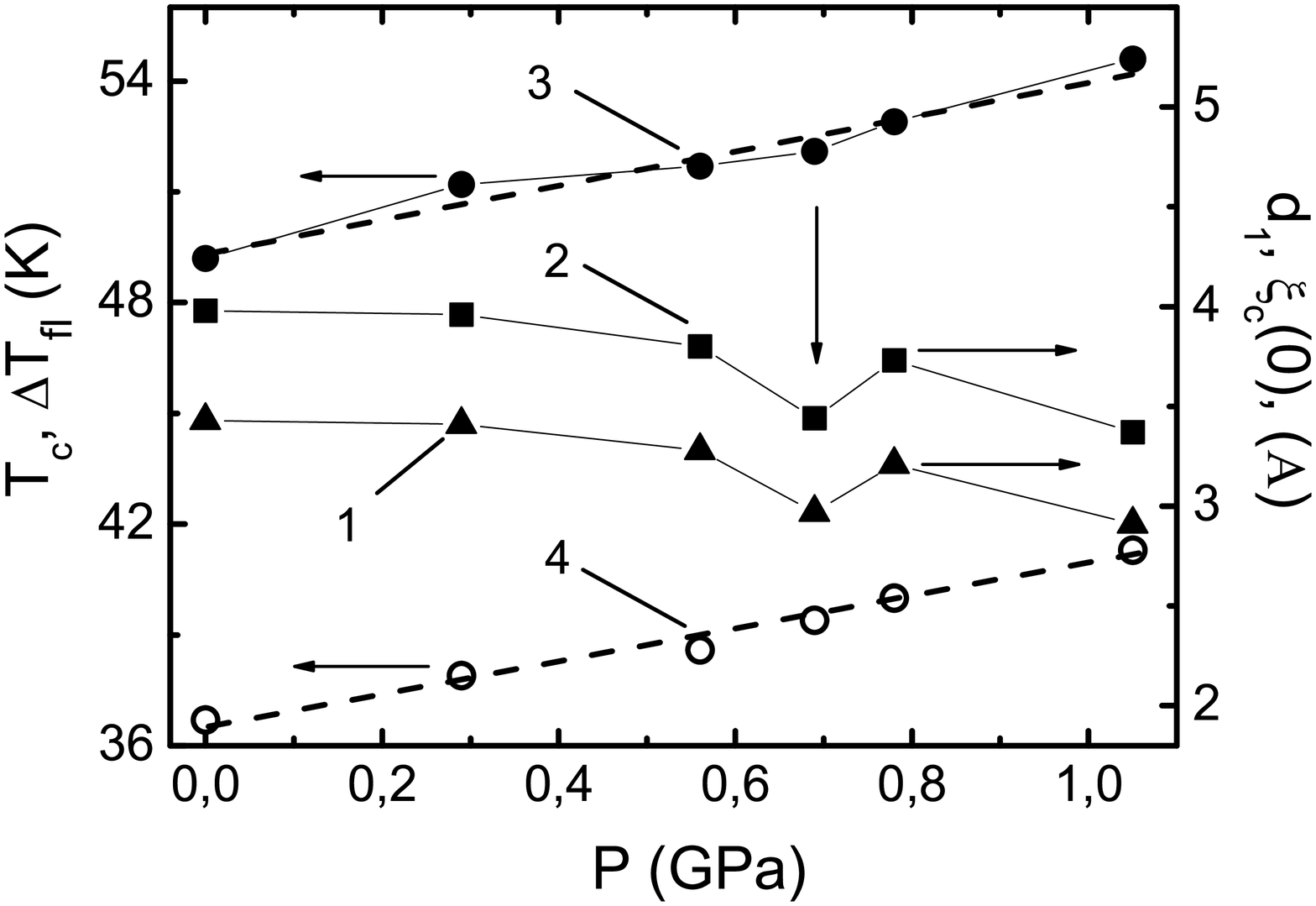}
\caption{Pressure dependence of $\xi_c(0)$ (curve 1), $d_1$ (curve 2), $T_c$ (curve 3) and $\Delta\,T_{fl}$ (curve 4).
 Solid lines are guides for the eye. Dashed lines are the lines of least-squares fit. Vertical arrow designates the peculiarity at $P \simeq 0.7\,GPa$.}
\end{center}
\end{figure}

\begin{figure}[b]
\begin{center}
\includegraphics[width=.52\textwidth]{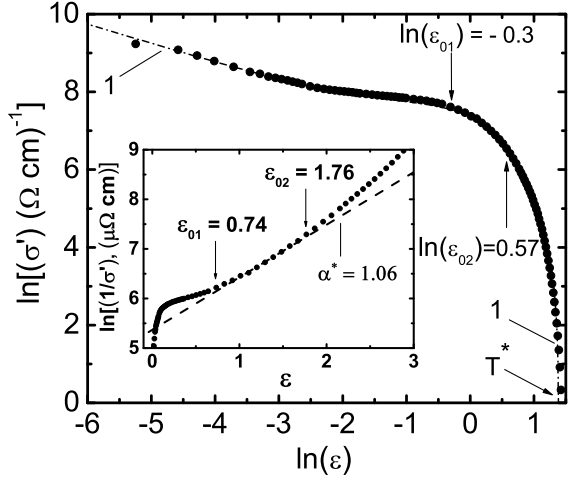}
\caption{$ln\sigma'$ vs $ln\varepsilon$ (dots) plotted in the whole temperature range from $T^*$ down to $T_c^{mf}$.
The dash-dotted curve (1) is fit to the data with Eq. (11).
Insert: $ln\sigma'^{-1}$ as a function of $\varepsilon$. Dashed line indicates the linear part of the curve between
$\varepsilon_{01}\simeq0.74$ and $\varepsilon_{02}\simeq1.76$. Corresponding $ln\varepsilon_{01}\simeq-0.3$ and $ln\varepsilon_{02}\simeq0.57$ are marked by the arrows at the main panel.
The slope $\alpha^*$=1.06 determines the parameter $\varepsilon_{c0}^* = 1/\alpha^*=0.94$.  P=0~GPa.}
\end{center}
\end{figure}

Importantly, the similar set of $\ln \sigma'$ vs $\ln \varepsilon$, as shown in Fig. 5, was obtained within our analysis for all applied pressure values used in the experiment.
The data demonstrate a very good fit with both, the 3D AL and the 2D MT theories in the whole temperature region of interest, in perfect agreement with the above considerations.
Somewhat surprisingly, despite of the pronounced decrease of the $\rho(P)$ (Figs. 1 and 2) the value of $\sigma'(T)$ is found to be nearly pressure-independent (Fig. 5).
Moreover, at all pressures the range of the wave function stiffness, or the range of the SC fluctuations, is restricted by $ln\varepsilon_{01} = -0.3\pm 0.01$ (Fig. 5).
At the same time, the crossover temperature $T_0$ is slightly shifted towards $T_c$ (Fig.5, b) suggesting an increase of the range of the 2D MT fluctuations.
As a result, both $\xi_c(0)$ and $d_1$ smoothly decrease with pressure (Table I) but have noticeable peculiarity again at about 0.7 GPa (Fig. 6, curves 1 and 2, respectively).
It is rather tempting to connect the observed peculiarity with $T_c$, which demonstrates a hint on the similar peculiarity at $P \simeq 0.7\,GPa$ (Fig. 6, curve 3).
But, on the one hand, the same increase of $T_c$ at P=0.29 GPa produces no effect on $\xi_c(0)$ and $d_1$  (Fig. 6).
On the other hand, if $T_c$ decreases, $\xi_c(0)$ has to increase (see Table I) in contrast with experiment.

It should be also noted that the range of FLC is located below 100 K.
At the same time, $\rho (P)$ measured at $100~K$ (Fig. 2, curve 2) demonstrates no visible peculiarity suggesting that the applied pressure is properly evaluated.
Thus, the observed peculiarity can likely be attributed to the specific reaction of the electronic subsystem of the studied single crystal on the applied pressure.
Apart from $T_c$ (Fig. 6, curve 3) all other characterictic FLC temperatures are also found to increas with pressure (Table I).
Moreover, the increase is mainly linear.
Also shown in Fig. 6 is $\Delta T_{fl}=T_{01}-T_G$ (curve 4) which determines the range of SC fluctuations above $T_c$ \cite{EK,Cor,Tal}.
The found depenmdence $\Delta T_{fl}(P)$ looks rather linear without any pronounced peculiarities.
Both, $T_{01}$ and $T_G$ also increase with increase of P, but $T_{01}$ increases a little bit faster, see Table I.
Thus, pressure increases the range of the SC fluctuations in which the stiffness of the wave function of the SC order parameter $\Delta$ is maintained \cite{EK,Cho,Cor,DTS}.
It is worthy to emphasize that both, $\Delta T_{fl}$ and $T_{01}$ are rather large, which is typical for the slightly doped cuprates \cite{EK,Cor,Tal,S6}.

\indent The obtained results allows one to consider the pressure effects on the extent of the resistive transition and the critical fluctuation regime \cite{S4,Wa,Pur,Sh}..
As can be seen from Fig. 6 and Table I, both $\xi_c(0)(P)$ and $d_1(P)$ are found to decrease at the similar rate with pressure.
The total decrease is about 15\% in both cases.
It also means that the coupling constant $\alpha$ by Eq. (6) or the coupling strength $J=[\xi_c(0)/d]^2$ for the neighboring  $CuO_2$ planes \cite{HL,Sh,Pur} is nearly pressure-independent.
This is in contrast with the HoBCO single crystals \cite{S4}, for which 9.4\% increase of the $\xi_c(0)(P)$ was found at almost constant d, resulting in increase of both, $\alpha$ given by Eq. (6) and $J=[\xi_c(0)/d]^2$.
However, like in HoBCO single crystals, the range of critical fluctuation, $\Delta\,T_{cr}=T_G - T_c$, noticeably increases from 1.5 K at P=0\,GPa up to 2.7 K at P=1.05~GPa (Fig. 4) resulting in the corresponding decrease of the 3D AL fluctuation region, $T_0$ - $T_G$, from 3.7~K down to 2.6~K (Figs. 3 and 4).
At the same time, in HoBCO the rate $d\Delta\,T_{cr}/dP$ is a factor of $\simeq 5$ larger, namely $5.6~KGPa^{-1}$ versus $1.14~KGPa^{-1}$.
As mentioned above, under pressure the width of the resistive transition, $\Delta\,T_c=T_c(0.9\rho'_N)-T_c(0.1\rho'_N)$ (Fig. 3), also increases but not so pronounced as found for HoBCO  \cite{S4}.
It is worthy to note that both, $T_G(P)$ and $\Delta\,T_{cr}(P)$ demonstrate the expected peculiarity at P=0.7\,GPa (Table II).

Having determined $T_G$ and $T_c^{mf}$ for each applied pressure one can calculate the Ginzburg number defined as $Gi=(T_G-T_c^{mf})/T_c^{mf}$.
Figs. 4a and b show that Gi also increases by about 20\% when P increases from 0  to P=1.05~GPa (see also Table II).
Together with the increase of $\Delta\,T_c$ (Fig. 3) and $\Delta\,T_{cr}$ (Fig. 4) it implies that the genuine critical fluctuations are somewhat enhanced when the pressure is applied.
It appears somewhat surprising that the pressure seems to improve the sample structure \cite{S4}.
But similar results have been obtained for OD $YBa_2Cu_3O_{7-\delta}$ \cite{Pur} as well as for HoBCO \cite{S4,S5}.
According to the anisotropic GL theory, the Ginzburg number is defined as \cite{Kap,Sch}
\begin{equation}
Gi = \alpha_1\biggl(\frac{k_B}{\Delta\,c\,\xi_c(0)\,\xi_{ab}(0)^2}\biggr)^2
\label{Gi}
\end{equation}
where $\alpha_1$ is a constant of the order of $10^{-3}$ and $\Delta\,c$ is the jump of the
specific heat at $T_c$.
According to the microscopic theory \cite{Sch}, $\Delta\,c \sim T_cN(0)$,
where $N(0)$ is the single-particle DOS at the Fermi level.
$\Delta\,c$ is expected to be weakly P-dependent in this range since $N(0)$, as deduced from the Pauli susceptibility
above $T_c$, is rather insensible to pressure in HTSCs \cite{Sc}.
The analysis of the FLC amplitude points to a 15\% decrease of $\xi_c(0)$ under pressure.
Together with the supposed corresponding decrease of $\xi_{ab}(0)$ it can completely provide the observed increase of Gi.

\begin{table}[tbp]
\caption [ ]

Parameters of the $YBa_2Cu_3O_{6.5}$ single crystal.
\centering
\begin{tabular}{||l|c|c|c|c|c|c|c|c|c|c||}
\hline\hline
P  & $\rho(100K)$ & $T_c$ & $T_c^{mf}$ & $T_{01}$ & $T_G$ & $\Delta\,T_{fl}$ & $d_1$ & $\xi_c(0)$\\ [0.5ex]
(GPa) & $(\mu\Omega cm)$ & $(K)$ & $(K)$ & $(K)$ & $(K)$ & $(K)$ & $(\AA)$ & $(\AA)$\\ [0.5ex]
\hline
0 & 180.4 & 49.2 & 50.2 & 87.4 & 50.7 & 36.7 & 3.98 & 3.43 \\
\hline
0.29 & 169.1 & 51.2 & 52.1 & 90.7 & 52.8 & 37.9 & 3.96 & 3.41 \\
\hline
0.56 & 159.2 & 51.7 & 52.6 & 91.6 & 53.0 & 38.6 & 3.8 & 3.28 \\
\hline
0.69 & 155.6 & 52.1 & 54.3 &  94.5 & 55.1 & 39.4 & 3.44 & 2.97 \\
\hline
0.78 & 152.4 & 52.9 & 54.8 & 95.4 & 55.4 & 40.0 & 3.73 & 3.21 \\
\hline
1.05 & 144.4 & 54.6 & 56.6 & 98.6 & 57.3 & 41.3 & 3.37 & 2.91 \\
\hline\hline
\end{tabular}
\label{tab:sample-values}
\end{table}

\begin{table}[tbp]
\caption [ ]

Parameters of the $YBa_2Cu_3O_{6.5}$ single crystal.
\centering
\begin{tabular}{||l|c|c|c|c|c|c|c|c||}
\hline\hline
P  & Gi & $\Delta\,T_{cr}$ & $T^*$ & D* & $\Delta^*(T_c)$  & $\Delta^*_{max}$ & $T_{max}$ & $T_{pair}$\\ [0.5ex]
(GPa) &  & $(K)$ & $(K)$ &  & $(K)$ & $(K)$ & $(K)$ & $(K)$ \\ [0.5ex]
\hline
0 & 0.01 & 1.5 &  252 &  5 & 122.1 & 184.17 & 231.6 & 170\\
\hline
0.29 & 0.013 & 1.6 & 252 &  5.4 & 136.5 & 192.28 & 229.2 & 165\\
\hline
0.56 & 0.008 & 1.3 & 252 &  5.8 & 145.8 & 199.76 & 226.2 & 159\\
\hline
0.69 & 0.015 & 3.0 & 252 &  6.4 & 164.3 & 190.71 & 217.0 & 153\\
\hline
0.78 & 0.011 & 2.5 & 253 &  6.5 & 167.7 & 190.3 & 152.8 & 138\\
\hline
1.05 & 0.012 & 2.7 & 254 &  6.6 & 178.4 & 198.41 & 205.7 & 135\\
\hline\hline
\end{tabular}
\label{tab:param}
\end{table}

This appears reasonable since
the relative increase of both the determined Ginzburg number $Gi^*=Gi(P)/Gi(0)\approx 1.25$ and $\Delta\,T_{cr}=\Delta\,T_{cr}(P)-\Delta\,T_{cr}(0)\approx 1.2~K$ is in good agreement with the corresponding parameters obtained for OD $YBa_2Cu_3O_{7-\delta}$  single crystals under the same pressure ($Gi^*\approx 1.25 $ and $\Delta\,T_{cr}\approx 0.4~K$) \cite{Pur}.
It should be also noted that both Gi(P) and $\Delta\,T_{cr}(P)$ demonstrate the expected peculiarity showing the anomalously large values just at $P\sim 0.7\,GPa $(Table II).
Naturally, we expected to find similar specific features in the PG behavior of our $YBa_2Cu_3O_{7-\delta}$  single crystals.\\

\indent {\bf B.\, Pseudogap analysis}\\

\indent The main subject of the present study is the pressure
effect on PG $\Delta^*$.
In the resistivity measurements of HTSCs (Fig. 1), PG becomes apparent through the downturn of the longitudinal resistivity $\rho(T)$  at $T\leq T^*$ from its linear behavior at higher temperatures above $T^*$ \cite{S2,Gus,SP,TS,P,Max}.
This results in appearance of the excess conductivity $\sigma'(T) = \sigma(T) - \sigma_N(T)$, as mentioned above (Eq. (1)).
If there were no processes in the HTSCs resulting in the PG opening at $T^*$, $\rho(T)$ would remain linear down to $\sim\, T_c$ \cite{Iye}.
Thus, the excess conductivity $\sigma'(T)$ emerges as a result of the PG opening.
Consequently, $\sigma'(T)$ has to contain information about the value and the temperature dependence of PG \cite{Kord,S1,L,Gus,Lang,ND}.

It is well established  now \cite{S1,S6,TS} that the conventional fluctuation theories
modified for the HTSCs by Hikami and Larkin (HL) \cite{HL} perfectly fit the experimental curves $\sigma'(T)$ but only up to approximately $T_{01} \simeq 110~K$.
Clearly, to attain information about the pseudogap one needs an equation which specifies the whole experimental curve from $T^*$ down to $T_c$ and contains PG in explicit form.
Besides, the dynamics of pair-creation $(1-T/T^*)$ and pair-breaking (exp$(-\Delta^*/T)$)
[see Eq.(11)] above $T_c$ must also be taken into account in order to correctly describe experiment \cite{S1,S2}.
Due to the absence of a complete fundamental theory, the equation for $\sigma'(\varepsilon)$ has been proposed in Ref. \cite{S2} with respect to the local pairs:
\begin{equation}
\sigma '(\varepsilon) = \frac{e^2\,A_4\,\left(1 -
\frac{T}{T^*}\right)\,\left(exp\left(-\frac{\Delta^*}{T}\right)\right)}{(16\,\hbar\,\xi_c(0)\,
\sqrt{2\,\varepsilon_{c0}^*\,\sinh(2\,\varepsilon\,/\,\varepsilon_{c0}^*})}.
\label{sigma-eps}
\end{equation}

Here $A_4$ is a numerical factor which has the  meaning of the C-factor in the FLC theory \cite{S1,S2,Oh}.
The values of $T^*$, $\varepsilon$ (Eq. (2)) and $\xi_c(0)$ by Eq. (4) have already been determined from the resistivity and the FLC analysis and are summarized in Table I.
The rest of parameters, such as the theoretical parameter $\varepsilon^*_{c0}$ \cite{LL}, the coefficient $A_4$, and $\Delta^*(T_c)$ \cite{St}, can be directly derived from the LP model analysis now \cite{S1,S2}.
In the range $ln\varepsilon_{01}<ln\varepsilon<ln\varepsilon_{02}$ (Fig. 7)  or accordingly $\varepsilon_{01}<\varepsilon<\varepsilon_{02}$ ($87.4~K<T<139~K$) (insert in Fig. 7),
$\sigma'^{-1} \sim$ exp$(\varepsilon$) \cite{S1,S2,LL}.
This feature turns out to be the basic property of the majority of the HTSCs including FeAs-based superconductors \cite{S7}.
As a result, in this temperature interval ln($\sigma'^{-1}$) is a linear function of $\varepsilon$ with a slope $\alpha^*$=1.06 which determines parameter $\varepsilon_{c0}^* = 1/\alpha^*=0.94$ \cite{LL} for the sample Y0 (insert in Fig. 7).
The same graphs, but with $\alpha^*$ increasing up to 1.4 ($\varepsilon_{c0}^*$=0.71) at P=1.05~GPa, where obtained for all applied pressures.
This allowed us to get reliable values of $\varepsilon_{c0}^*$ which are found to noticeably affect the shape of the theoretical curves displayed in Figs. 7 and 8.
\begin{figure}[t]
\begin{center}
\includegraphics[width=.58\textwidth]{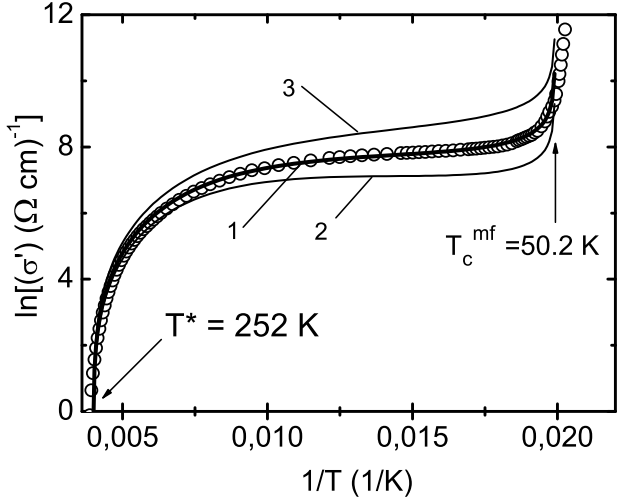}
\caption{$ln\sigma'$ vs 1/T (circles) plotted in the whole temperature range from $T^*$ down to $T_c^{mf}$ at P=0.
The solid curves are fits to the data with Eq. (10).
The best fit is obtained when Eq. (10) is calculated for $\Delta^*(T_c)$=122.1~K ($D^*=2\Delta^*(T_c^{mf})/k_B\,T_c= 5.0$ (curve 1)). Curves 2 and 3 corresponds to D*=6 and 4, respectively.}
\end{center}
\end{figure}

To find $A_4$, we calculate  $\sigma'(\varepsilon)$ using Eq. (10) and fit the experimental data in the range of 3D AL
fluctuations near $T_c$ (Fig. 7) where $ln\sigma'(ln\varepsilon)$ is a linear function of the reduced temperature $\varepsilon$ with the slope $\lambda$ = $-$1/2.
Besides, $\Delta^*(T_c^{mf}) = \Delta_0(0)$ is assumed \cite{S2,St}.
To estimate $\Delta^*(T_c^{mf})$, which we use in Eq. (10), we plot $ln\sigma'$ as a function of 1/T \cite{S2,Pr} (Fig. 8, circles).
In this case the slope of the theoretical curve by Eq. (10) turns out to be very sensitive to the value $\Delta^*(T_{c})$ \cite{S1,S2}.
The best fit (solid curve 1 in the figure) is obtained when $D^*=2\Delta^*(T_c^{mf})/k_B\,T_c= 5.0\pm0.1$, which is typical for the underdoped YBCO cuprates \cite{S1,S6,Wan,Za}.
With all found parameters Eq. (10) perfectly describes the experimental $ln\sigma'(1/T)$ now (Fig. 7, dashed curve 1).
Similar graphs were revealed at all pressures applied.
In all cases a very good fit is obtained allowing us to get reliable values of D* at all pressure values (Table II).
Note that the found $D^*=2\Delta^*(T_c^{mf})/k_B\,T_c$=5 (P=0) and $D^*$=6.6 (P=1.05\,GPa) correspond to the strong coupling limit being typical for HTSCs in contrast to the BCS weak coupling limit ($2\Delta_0/k_B\,T_c^{BCS} \approx 4.28$ established for the d-wave superconductors \cite{Ino,Fish}).
The found values are in good agreement with
results of the mentioned above $\mu SR$ experiment \cite {Mai} in which a similar increase of the SC gap $\Delta_0$ as well as of the BCS ratio $2\Delta_0/k_B\,T_c$ is reported at the same pressure.

Solving Eq. (10) for the pseudogap $\Delta^*(T)$ one can readily obtain
\begin{equation}
\Delta^*(T) = T\,ln\frac{e^2\,A_4\,(1 - \frac{T}{T^*})}{\sigma '(T)\,16\,\hbar\,\xi_c(0)\,\sqrt{2\,\varepsilon_{c0}^*\,\sinh(2\,\varepsilon\,/\,\varepsilon_{c0}^*)}}.
\label{delta-t}
\end{equation}
Here $\sigma'$(T) is the experimentally measured excess conductivity in the whole temperature interval from $T^*$ down to $T_c^{mf}$.
The fact that $\sigma'(T)$ is perfectly described by Eq. (10) (Figs. 7 and 8) allows one to conclude that Eq. (11) yields reliable both, the magnitude and the temperature dependence of the PG now.
Fig.~9 displays the results of the PG analysis for samples Y0 and Y6.
The bottom curve 1 (dots) is computed at P=0 using Eq.(11) with the following set of parameters derived from experiment:
$T^* = 252$\,K, $T_c^{mf} = 50.2$\,K, $\xi_c(0) = 3.43$\AA, $\varepsilon_{c0}^* = 0.94$, $A_4 = 55$ and $\Delta^*(T_c)/k_B$ = 122.1\,K (Tables I and II).
The upper curve 2 (semicircles) is computed at P=1.05\,GPa with the following set of parameters:
$T^* = 254$\,K, $T_c^{mf} = 56.6$\,K, $\xi_c(0) = 2.91$\AA, $\varepsilon_{c0}^* = 0.71$, $A_4 = 100$ and $\Delta^*(T_c)/k_B$ = 178.4\,K (Tables I and II).
As can be seen in the figure, both curves look rather similar, especially in the range of low temperatures.
At the same time, a pronounced increase of $\Delta^*$ with increase of P is observed.
The values of $\Delta^*(P)$ for the slightly doped samples investigated in this work
are summarized in Table II.
The revealed increase of PG under pressure is explicitly observed for the first time and
represents the main result of our study.
The details of the PG behavior under hydrostatic pressure are analyzed next.
\begin{figure}[b]
\begin{center}
\includegraphics[width=.60\textwidth]{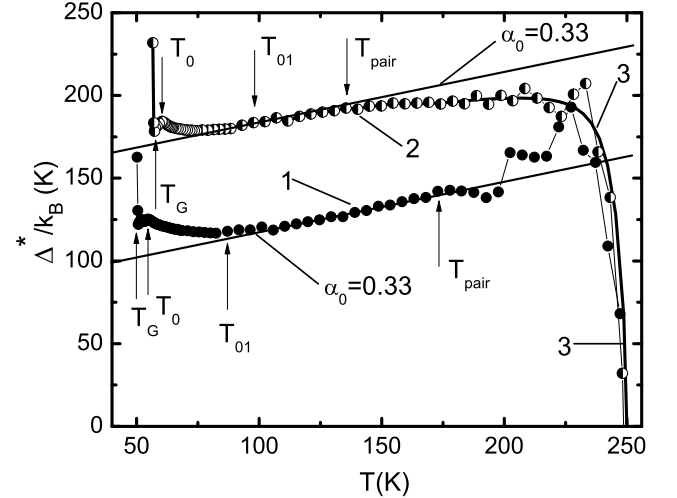}
\caption{Temperature dependence of the pseudogap $\Delta^*$ of the $YBa_2Cu_3O_{6.5}$ single crystal at zero pressure (curve 1, dots) and P=1.05~GPa (curve 2, semicircles). The data were analyzed using Eq. (11).
Solid curve 3 indicates the result of such analysis performed at P=1.05\,GPa but using the resistivity curve polynomially fitted down to $\approx 160~K$.
The solid lines indicate the linear $\Delta^*(T)$ dependencies below $T_{pair}$, whose slope $\alpha_0 \simeq0.33$ is found to be pressure-independent, indicating the noticeable increase of $\Delta^*$  with pressure.
All temperatures marked by arrows in the figure are discussed in the text.}
\end{center}
\end{figure}

Fig. 9 shows that each typical feature of the $\Delta^*(T)$ curve occurs at the corresponding characteristic temperature.
Indeed, below $T_{pair}$ $\Delta^*(T)$ appears to be linear down to $T_{01}$.
The linearity is designated by the straight solid lines with the slope  $\alpha_0=0.33\pm0.01$ which
is found to be pressure-independent (see also Fig. 10).
In accordance with the LP model, above $T_{pair}$ the LPs have to exist mostly in the form of SBBs.
Below $T_{pair}$ they have to transform into FCP.
Thus, $T_{pair}$ separates both regimes \cite{S1,ST,L,Gus}.
Here $T_{pair}$ is introduced as a temperature at which $\Delta^*(T)$ turns down from its linear behavior with increase of T.
Below $T_{01}$ PG increases gradually likely due to the formation of SC fluctuations, showing maximum at $T_0$.
Below $T_0$ it rapidly decreases down to $T_G$, which is determined as a leftmost temperature with the reliable $\Delta^*$  value.
Below $T_G$ $\Delta^*$ increases irregularly because this is the region of critical fluctuations where the LP model does not work.
Thus, the LP model approach allows one to get precise values of $T_G(P)$ and, hence, reliable values of $\Delta^*(T_c^{mf}) \simeq \Delta^*(T_G)$ summarized in Tables I and II.
Evidently, $\Delta^*$ noticeably increases with increasing pressure as mentioned above.
\begin{figure}[t]
\begin{center}
\includegraphics[width=.60\textwidth]{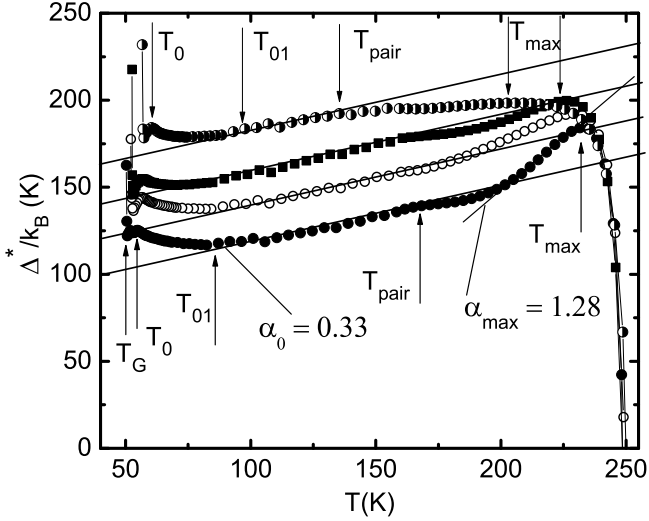}
\caption{Temperature dependence of the pseudogap $\Delta^*$ of $YBa_2Cu_3O_{6.5}$ single crystal at different applied hydrostatic pressures, from bottom to top: dots - P=0\,GPa; circles - P=0.29\,GPa; squares - P=0.56\,GPa, semicircles - P=1.05\,GPa.
The curves computed at P=6.9\,GPa and P=7.8\,Gpa are not shown to simplify reading the data.
The data were analyzed with Eq. (11) but using resistivity curves polynomially fitted down to $\sim 160\,K$  (see  text for details).
Both $\Delta^*$ and $2\Delta^*(T_c^{mf})/k_B\,T_c$ increase with increasing pressure at the same rate
$dln\Delta^*/dP\approx 0.36\,GPa^{-1}$.
The solid lines indicate the linear $\Delta^*(T)$ dependencies below $T_{pair}$ with the pressure-independent slope $\alpha \simeq0.33$.
All temperatures marked by arrows in the figure are discussed in the text.}
\end{center}
\end{figure}

In many experiments \cite{S2,LL,Pr} it was revealed that $\Delta^*(T)$ determined by Eq. (11) is very sensitive to
any uncertainties in the $\sigma'(T)$ determination which stems from the spread in the resistivity data.
It results in pronounced jumps of calculated $\Delta^*(T)$ at high temperatures near $T^*$ (Fig. 9), where $\sigma'(T)$ changes at a very high rate (see Fig. 7), which have no intrinsic physical meaning.
To avoid these jumps all resistivity curves were fitted by a polynomial down to $\sim 160\,K$ (e.g. see Fig. 1 curve 2).
Solid curve 3 in Fig. 9 indicates the result of such an approach based on the analysis of the resistivity curve fitted at P=1.05\,GPa.
As is evident from the figure, the polynomial fit provides a precise description of the raw data.

Analogous $\Delta^*(T)$ curves calculated at P=0, 0.29, 0.56 and 1.05\,GPa using the same fitting procedure are plotted in Fig. 10.
The curves calculated at P=6.9 and 7.8 \,GPa are not shown to simplify reading the data.
The calculations were performed using Eq. (11) with the corresponding sets of parameters determined in the above analysis (Tables I and II).
It should be emphasized that all parameters listed in the Tables, except $T_{max}$  and $\Delta^*_{max}(T_{max})$, where determined using FLC and the $\Delta^*(T)$ curves calculated with the raw resistivity data.
As can be observed in the figure, the only difference from the curves shown in Fig. 9 is the appearance of the expected high-temperature maximum at $T_{max}$ followed by the region of the linear $\Delta^*(T)$ behavior with the positive slope $\alpha_{max}=1.28\pm0.01$.
The maximum appears to be a typical feature of the SD single crystal PG behavior \cite{S4,S5}.
With increase of pressure, the slope $\alpha_{max}$ rapidly decreases and the maximum disappears already at $P\sim 0.7\,GPa$.
Simultaneously, $T_{max}$  as well as $T_{pair}$ decreases gradually with P, whereas the corresponding $\Delta_{max}(T_{max})$ and $\Delta^*(T_c^{mf})$ increase, but again with the expected peculiarity at $P\sim 0.7\,GPa$ as shown in Fig. 11, curve 1 and curve 2, respectively.
Thus, the bulk of the sample parameters demonstrate the peculiarity at $P\sim 0.7\,GPa$ which appears to be the particular pressure in our case.
As a result, with increase of pressure the shape of the $\Delta^*(T)$ dependence changes noticeably in the range of high T.
But it remains almost pressure-independent below $T_{pair}$.
Eventually, at P=1.05\,GPa the $\Delta^*(T)$ curve acquires the specific shape with $T_{pair}\approx (133\pm2)~K$ being typical for SD YBCO films at P=0 \cite{S1,S2}.
This suggests a strong influence of pressure on the lattice dynamics \cite{Mai} especially in the high-temperature region.
\begin{figure}[b]
\begin{center}
\includegraphics[width=.54\textwidth]{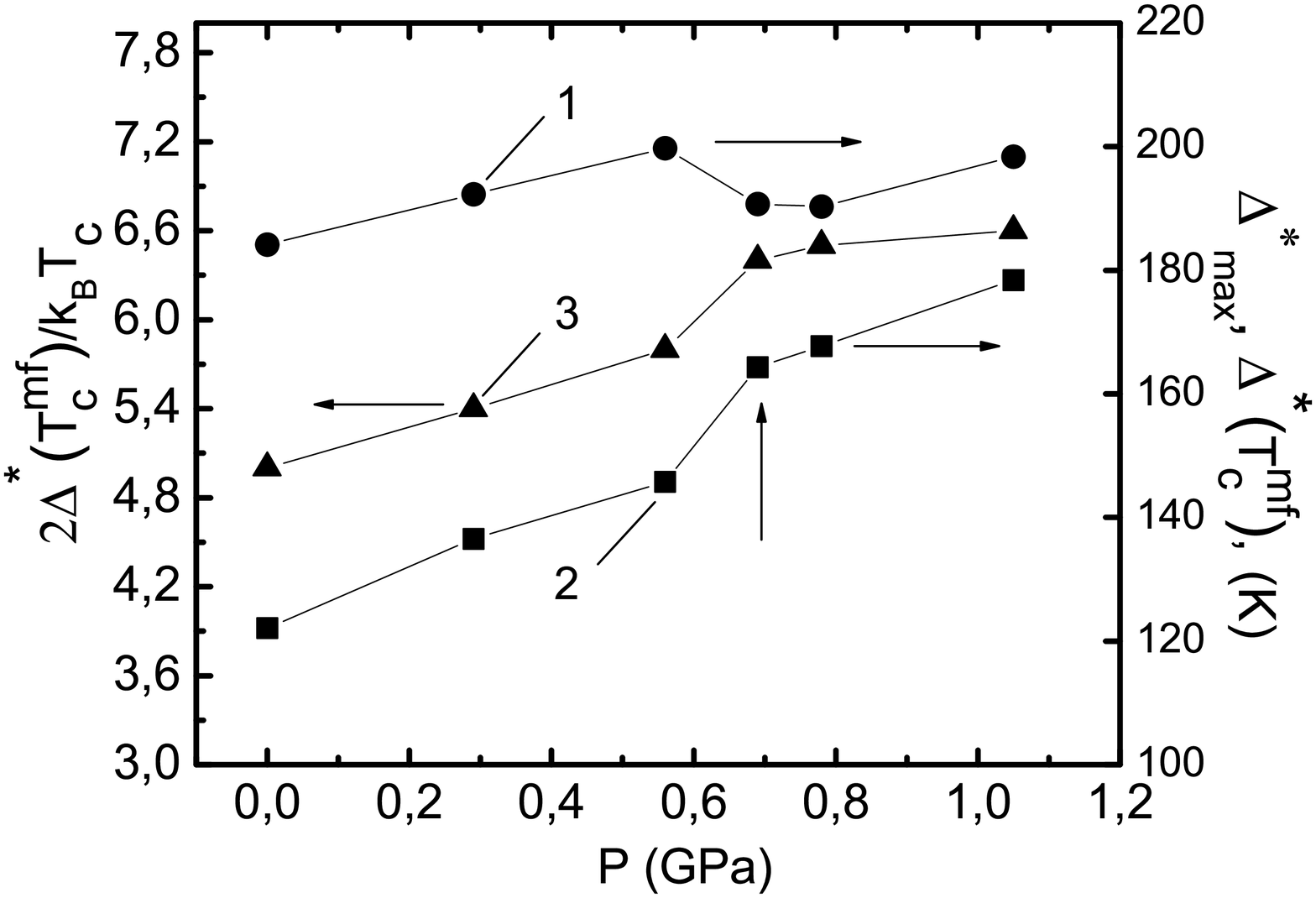}
\caption{Pressure dependence of $\Delta^*_{max}$ (curve 1), $\Delta^*(T_c^{mf})$ (curve 2)
and $D^*=2\Delta^*(T_c^{mf})/k_B\,T_c$ (curve 3).
Solid lines are guides for the eye. Vertical arrow marks the peculiarity at $P\sim 0.7\,GPa$}.
\end{center}
\end{figure}

Also plotted in Fig. 11 is $D^*=2\Delta^*(T_c^{mf})/k_B\,T_c$ (curve 3, triangles).
As expected, D* and $\Delta^*(T_c^{mf})$ (curve 2, squares) both demonstrate identical pressure dependences.
It is important to emphasize that $2\Delta^*(T_c^{mf})/k_B\,T_c$ has been obtained from the fundamentally different approach, namely from fitting of the measured excess conductivity to Eq. (10), as illustrated in Fig. 8.
Thus, the different approaches give the same result which indicates that PG increases with pressure
at a rate $dln\Delta^*/dP\approx 0.36\,GPa^{-1}$.
This value is a factor of $\simeq 3.3$ larger than that reported using tunneling spectroscopy of Ag-Bi2223 point contacts \cite{DT} but again in good agreement with the results
of the mentioned above $\mu SR$ experiment on the SD polycrystalline $YBa_2Cu_3O_{7-\delta}$ \cite{Mai}.

Thus, for slightly doped YBCO single crystals both $\Delta^*$ and the BCS ratio $D^*=2\Delta^*(T_c^{mf})/k_BT_c$
increase upon increasing applied pressure.
This suggests an increase of the coupling strength with increasing pressure.
Strictly speaking, the increase of PG, as well as the SC gap \cite{Mai,DT} in the HTSCs under hydrostatic pressure remains even less comprehensible than the corresponding increase of $T_c$.
In fact, the pressure dependence of the SC transition temperature is determined by two
mechanisms.
First, it is the pressure induced charge transfer to $CuO_2$ planes, $\Delta n_h$, as mentioned above.
The second mechanism consists in the possible increase of the pairing interaction $V_{eff}$ which depends on pressure.
For underdoped cuprates the former mechanism
dominates in the pressure effect on $T_c$ (see Ref \cite{Mai} and references therein).
It is well known that in YBCO the unique proximity between the Cu and O states is realized \cite{L,Im}.
As a result, the band structure of the cuprate HTSCs is determined by the strongly correlated electron motion on the
$Cu(3d)$ orbital which is likely under the influence of the $O(2p)$ one.
Hydrostatic pressure can very likely affect the interaction.
The possible increase of $V_{eff}$ under pressure could result in increase of both $\Delta_0(P)$ and $\Delta^*(P)$.
However, it is not clear whether this mechanism strong enough to provide the observed rather large increase of both, the SC gap and the pseudogap.

Details of the electron-phonon interaction in HTSCs, which are known to be rather specific \cite{Max,Zu}, were thoroughly analyzed in Ref.\cite{DT} by means of tunneling spectroscopy of Ag-Bi2223 point contacts under hydrostatic pressure.
It was shown that the SC gap $\Delta_0$ increases with increasing pressure at a rate $dln\Delta_0/dP\simeq\,0.1\,GPa^{-1}$ resulting in the corresponding increase of D*.
Simultaneously, the phonon spectrum was found to be noticeably shifted towards low energies.
The anomalous softening of the phonon frequencies under pressure is believed to be the result of the specific electron-phonon interaction in Bi2223 arising from the electron hopping between the conducting $CuO_2$ planes and is concluded to cause the observed increase of the SC gap (see Ref.\cite{DT} and references therein).
Unfortunately, there is lack of such experiments for YBCO.
As a result, the physics behind the pronounced increase of the pseudogap $\Delta^*$ observed in the present study under hydrostatic pressure as well as the corresponding increase of the superconducting gap $\Delta_0$ reported in Ref. \cite{Mai} still remains uncertain.

\indent {\bf IV\, Conclusion}\\

The pressure dependence of the resistivity, excess conductivity $\sigma'(T)$
and pseudogap (PG)\, $\Delta^*(T)$ of slightly doped single crystals of $YBa_2Cu_3O_{7-\delta}$ was studied within the local pair model.
It was expectedly found that with increasing pressure the sample resistivity $\rho$ decreases at a rate
$dln\rho(300\,K)/dP = (- 19\pm 0,2)\%~GPa^{-1}$
whereas the critical temperature $T_c$ increases at a rate $dT_c/dP = +5.1~KGPa^{-1}$, which both are in good agreement with those obtained for the YBCO compounds by different experimental technique.
Pressure is believed to stimulate the processes of the charge carriers redistribution resulting in increase of $n_f$ in the conducting CuO$_2$ planes, that should lead to the observed reduction of $\rho$ as well as to the increase of $T_c$.
Simultaneously, the noticeable decrease of the distance between the conducting $CuO_2$ planes $d_1$ with pressure was observed.
It could lead to a modification of the pairing interaction $V_{eff}$ by pressure, which, in turn, also can lead to increase $T_c$.
Independently on pressure near $T_c$, $\sigma'(T)$ is well described by the Aslamasov-Larkin and
Hikami-Larkin fluctuation theories demonstrating a 3D-2D crossover with increase
of temperature.
The crossover temperature $T_0$ determines the coherence length along the c-axis $\xi_c(0)\simeq(3.43\pm0.01)$\AA\
at P=0\,GPa.
The revealed value of $\xi_c(0)$ is typical for the slightly doped cuprates and it is found to decrease with P.
The rest of the sample parameters also change with increasing pressure, demonstrating a noticeable peculiarity at
$P \simeq0.7\,GPa$, suggesting a strong influence of pressure on the lattice dynamics likely due to the pressure effect on the pairing interaction in the cuprates.

The same conclusion arises from the observation that pressure noticeably changes the shape of the $\Delta^*(T)$ curve at high temperatures above $T_{pair}$ but leaves it almost invariant at $T<T_{pair}$.
The pseudogap $\Delta^*$ and the BCS ratio $D^*=2\Delta(T_c^{mf})/k_B\,T_c$ both increase with increasing applied hydrostatic pressure at a rate $dln\Delta^*/dP\approx 0.36\,GPa^{-1}$, implying an increase of the coupling strength in the curates with pressure.
The explicit modification of the temperature dependence of PG in the slightly doped HTSCs with increasing pressure is observed for the first time.
The found rate of the PG modification is a factor of $\simeq 3.3$ larger than that observed by pressure experiments in OD polycrystalline Bi2223 using a tunneling technique.
In Bi compounds the enhancement of both, SC gap $\Delta_0$ and $2\Delta_0(0)/k_B\,T_c$ was attributed to
the specific electron-phonon interaction in Bi2223 arising from possible electron hopping between the conducting
$CuO_2$ planes.
It seems to be rather tempting to ascribe the observed increase of the PG in $YB_2Cu_3O_{6.5}$ single crystal to the similar softening of the phonon spectra under pressure.
However, there is lack of such experiments with YBCO, and the physics behind the observed PG increase under hydrostatic pressure still remains uncertain, thus demanding a further study.\\



\begin{thebibliography}{85}
%
\bibitem{Kord} A. A. Kordyuk, arXiv:1501.04154v1 [cond-mat.suprcon] (2015).
\bibitem{S1} A. L. Solovjov, Superconductors - Materials, Properties and Applications. Chapter 7: Pseudogap and local pairs in high-Tc superconductors, InTech, Rijeka, 137 (2012).
\bibitem{PB} R. Peters, and J. Bauer, arXiv:1503.03075v1 [cond-mat.suprcon] (2015).
\bibitem{M} N. F. Mott, Rev. Mod. Phys. {\bf 40}, 677 (1968).
\bibitem{M2} N.F. Mott, Metal–Insulator Transition, World Scientific, London, 1974.
\bibitem{Al} H. Alloul, T. Ohno, and P. Mendels, Phys. Rev. Lett. {\bf 63}, 1700 (1989).
\bibitem{LP} E. M. Lifshitz and L. P. Pitaevski, Statistical Physics, {\bf vol. 2}, Moscow: Nauka, 1978.
\bibitem{Kon} Takeshi Kondo, A. D. Palczewski, Y. Hamay et al. arXiv:1208.3448v1 (2012).
\bibitem{Tel} L. Taillefer, Annu. Rev. Condens.Matter Phys. {\bf 1}, 51 (2010).
\bibitem{Gab} A. M. Gabovich and A. I. Voitenko, Phys. Rev. B, {\bf 80}, 224501 (1-9) (2009).
\bibitem{Nor} M.R. Norman, in: Novel Superfluids, vol. 2, K.H. Bennemann and J.B. Ketterson
(eds.), Oxford University Press (2013).
\bibitem{Ber} C. Berthod, Y. Fasano, I. Maggio-Aprile, A. Piriou, E. Giannini, G. Levy de Castro,
Q. Fischer, Phys. Rev. B {\bf 88}, 014528 (2013).
\bibitem{EK} V. J. Emery and S. A. Kivelson, Nature, {\bf 374}, 434 (1995).
\bibitem{Cho} H-Y. Choi, Y. Bang, and D. K. Campbell, Phys. Rev. B {\bf 61}, 9748 (2000).
\bibitem{Yaz} A Yazdani, J. Phys.: Condens. Matter {\bf 21}, 164214 (9pp) (2009).
\bibitem{Mis} Vivek Mishra, U. Chatterjee, J. C. Campuzano, and M. R. Norman, Nature Phys. Lett. {\bf 10}, 357 (2014).
\bibitem{Tch} O. Tchernyshyov, Phys. Rev. B {\bf 56},  3372 (1997).
\bibitem{Kag} R. Combescot, X. Leyronas, and M. Yu. Kagan, Phys. Rev. A {\bf 73}, 023618 (2006).
\bibitem{L} V.M. Loktev, Low Temp. Phys. {\bf 22}, 488 (1996).
\bibitem{H} R. Haussmann, Phys. Rev. B, {\bf 49}, 12975 (1994).
\bibitem{Eng} J.R. Engelbrecht, A. Nazarenko, M. Randeria, and E. Dagotto, Phys. Rev. B {\bf 57}, 13406 (1998).
\bibitem{Sug} J. Sugawara, H. Iwasaki, N. Kabayashi, H. Yamane, and T. Hirai. Phys. Rev. B {\bf 46}, 14818 (1992).
\bibitem{WzK} Winzer K. and Kumm G. Z. Phys. B.- Condensed Matter,{\bf 82}, 317 (1991).
\bibitem{Kon2} T. Kondo, Y. Hamaya, Ari D. Palczewski, T. Takeuchi, J. S.Wen, Z. J. Xu, G. Gu, J. Schmalian and A. Kaminski, Nature Phys Lett. {\bf 7}, 21 (2011).
\bibitem{DeGen} P. G. De Gennes, Superconductivity of metals and alloys (W. A. Benjamin, INC., New York - Amsterdam, 1966), p. 280.
\bibitem{ST} A. L. Solovjov, M. A. Tkachenko, Metallofiz. Noveishie Tekhnol. {|bf 35}, 19 (2013), arXiv:1112.3812v1 [cond-mat.supr-con] (2012).
\bibitem{S2} A. L. Solovjov, V. M. Dmitriev. Low Temp. Phys. {\bf 32}, 99 (2006).
\bibitem{Cor} J. Corson, R. Mallozzi and J. Orenstein , J.N. Eckstein, I. Bozovic. Nature,  {\bf 398}, 221 (1999).
\bibitem{Tal} J. L. Tallon, F. Barber, J. G. Storey and J. W. Loram, Phys.Rev. B {\bf 87}, 140508(R) (2013).
\bibitem{DTS} A.I. D'yachenko, V. Yu. Tarenkov, S. L. Sidorov, V. N. Varyukhin, and A. L. Solovjov, Low. Temp. Phys. {\bf 39}, 323 (2013).
\bibitem{Gus} V. P. Gusynin, V. M. Loktev, S. G. Sharapov. ZETF Lett. {\bf 65}, 170 (1997).
\bibitem{Liu} H.J. Liu, Q. Wang, G.A. Saunders, D.P. Almond, B. Chapman, K. Kitahama, Phys.
Rev. B {\bf 51},  9167 (1995).
\bibitem{Wan} Q. Wang, G.A. Saunders, H.J. Liu, M.S. Acres, D.P. Almond, Phys. Rev. B {\bf 55}, 8529 (1997).
\bibitem{Fer} L.M. Ferreira, P. Pureur, H.A. Borges, P. Lejay, Phys. Rev. B {\bf 69}, 212505 (2004).
\bibitem{She} L.J. Shen, C.C. Lam, J.Q. Li, J. Feng, Y.S. Chen, H.M. Shao, Supercond. Sci. Technol.
{\bf 11}, 1277 (1998).
\bibitem{V1} R.V. Vovk, Z.F. Nazyrov, M.A. Obolenskii, I.L. Goulatis, A. Chroneos, V.M. Pinto
Simoes, Philos. Mag. {\bf 91}, 2291 (2011).
\bibitem{V2} R.V. Vovk, M.A. Obolenskii, Z.F. Nazyrov, I.L. Goulatis, A. Chroneos, V.M. Pinto
Simoes, J. Mater Sci.: Mater. Electron. {\bf 23}, 1255 (2012).
\bibitem{V3} R.V. Vovk, A.A. Zavgorodniy, M.A. Obolenskii, I.L. Goulatis, A. Chroneos, V.P.
Pinto Simoes, J. Mater. Sci. Mater. Electron. {\bf 22}, 20 (2011).
\bibitem{Chu} C.W. Chu, P.H. Hor, R.L. Meng, L. Gao, A.J. Huang, and Y.Q. Wang, Phys. Rev. Lett. {\bf 58},
405 (1988).
\bibitem{Fan} Y. Fang, D. Yazici, B. D. White, and M. B. Maple, arXiv:1507.03172v1 [cond-mat.supr-con], 2015.
\bibitem{S4} A. L. Solovjov, M. A. Tkachenko, R. V. Vovk , A. Chroneos, Physica C {\bf 501}, 24 (2014).
\bibitem{Cav} R. J. Cava,  Science {\bf 243} (4943), 656 (1990).
\bibitem{Ast} M. Asta, D. de Futaine, G. Ceder. E. Salomons, and M. Kraitchman, J. Less. Common Metals {\bf 168}, 39 (1991).
\bibitem{Kha} R. Khasanov, M. Bendele, A. Amato, K. Conder, H. Keller, H.-H. Klauss, H. Luetkens, and E. Pomjakushina,  arXive:0912.0471v1[cond-mat.supr-con], 2009.
\bibitem{DT} A.I. D'yachenko, V. Yu. Tarenkov, Phys. Techn. High Pres. {\bf 24}, 24 (2014).
\bibitem{S5} A. L. Solovjov, M. A. Tkachenko, R. V. Vovk , M. A. Obolenskii, Low. Temp. Phys. {\bf 37},
840 (2011).
\bibitem{V4} R. V. Vovk, M. A. Obolenskii, A. A. Zavgorodniy, I. L. Goulatis, V. I. Beletskii, A.
Chroneos, Physica C {\bf 469}, 203 (2009).
\bibitem{V5} R.V. Vovk, M.A. Obolenskii, A.A. Zavgorodniy, A.V. Bondarenko, I.L. Goulatis,
A.V. Samoilov, A.I. Chroneos, J. Alloys Comp. {\bf 453}, 69 (2008).
\bibitem{Tho} J.D. Thompson, Rev. Sci. Instrum. {\bf 55} 231 (1984).
\bibitem{Ito} T. Ito, K. Takenaka, and S. Uchida, Phys. Rev. Lett. {\bf 70}, 3995(1993).
\bibitem{Mo} B. Wuyts, V. V. Moshchalkov, and Y. Bruynseraede. Phys. Rev. B {\bf 53}, 9418 (1996).
\bibitem{An} Y. Ando, S. Komiya, K. Segawa, S. Ono, and Y. Kurita, Phys. Rev. Lett. {\bf 93}, 267001 (2004).
\bibitem{DeM} E. V. L. de Mello, M. T. D. Orlando, J. L. Gonzalez, E. S. Caixeiro, and E. Baggio-Saitovich, Phys. Rev. B, {\bf 66}, 092504 (2002).
\bibitem{Lang} W. Lang, G. Heine, P. Schwab, X. Z. Wang, and D. Bauerle,  Phys. Rev. B, {\bf 49}, 4209 (1994).
\bibitem{Mai} A. Maisuradze, A. Shengelaya, A. Amato, E. Pomjakushina, and H. Keller. Phys. Rev. B {\bf 84}, 184523 (2011).
\bibitem{SP} B.	P. Stojkovic, D. Pines, Phys. Rev. B {\bf 55}, 8576 (1997).
\bibitem{Oh}  B. Oh, K. Char, A. D. Kent, M. Naito, M. R. Beasley et al., Phys. Rev. B, {\bf 37}, 7861 (1988).
\bibitem{ND} R. K. Nkum and W. R. Datars, Phys. Rev. B, {\bf 44}, 12516 (1991).
\bibitem{GL1} V. L. Ginzburg, L. D. Landau, JETP, {\bf 20}, 1064 (1950).
\bibitem{HL} S. Hikami, A.I. Larkin, Mod. Phys. Lett. B, {\bf 2}, 693 (1988).
\bibitem{Beas} M. R. Beasley, Phisica B {\bf 148}, 191 (1987).
\bibitem{Xie} Y. B. Xie, Phys. Rev. B, {\bf 46}, 13997 (1992).
\bibitem{AL} L. G. Aslamazov and A. L. Larkin, Phys. Lett., {\bf 26A}, 238 (1968).
\bibitem{Mak} K. Maki,  Prog. Theor. Phys., {\bf 39}, 897 (1968).
\bibitem{Th} R. S. Thompson, Phys. Rev. B, {\bf 1}, 327 (1970).
\bibitem{S6}  A.L. Solovjov, H.-U. Habermeier, T. Haage, Low Temp. Phys., \textbf{28}, 22 (2002), and, \textbf {28}, 144 (2002).
\bibitem{Kap} A. Kapitulnik, M. R. Beasley, C. Castellani, and C. Di Castro,
Phys. Rev. B {\bf 37}, 537 (1988).
\bibitem{Chr} G.D. Chryssikos,  E.I. Kamitsos, J.A. Kapoutsis  et al. : Physica C \textbf {254}, 44 (1995).
\bibitem{Gosh} B.N. Goshchitskii, V.L. Kozhevnikov, M.V. Sadovskii, Int. J. Mod. Phys. {\bf B2}, 1331 (1988).
\bibitem{S7} A. L. Solovjov, V. N. Svetlov, V. B. Stepanov, S. L. Sidorov, V.Yu. Tarenkov, A. I.D'yachenko, and A.B.Agafonov, arXiv:1012.1252v [cond-mat,supr-con].
Twelfth Int. Conf. on Low Temp. Phys., Kyoto, 1970, edited by E. Kanda (Keigaku, Tokyo,
1971), p. 361.
\bibitem{Wa} Q. Wang, G. A. Saunders, H. J. Liu, M. S. Acres, and D. P. Almond, Phys. Rev. B {\bf 55}, 8529 (1997).
\bibitem{Pur} L. M. Ferreira, P. Pureur, H. A. Borges, and P. Lejay, Phys. Rev. B {\bf 69}, 212505 (2004).
\bibitem{Sh} L J Shen, C C Lam, J Q Li, J Feng, Y S Chen and H M Shao, Supercond. Sci. Technol. {\bf 11}, 1277 (1998).
\bibitem{Sch} T. Schneider and J. M. Singer, Phase Transition Approach to
High Temperature Superconductivity: Universal Properties of
Cuprate Superconductors (Imperial College Press, London, 2000).
\bibitem{Sc} J. S. Schilling and S. Klotz, in Physical Properties of High Temperature
Superconductors, edited by D. M. Ginsberg (World Scientific, Singapore, 1992), {\bf Vol. 3}, p. 59.
\bibitem{Iye} Y. Iye Transport properties of high Tc cuprates. Phys. Properties of High- Temp. Superconductors.  Ed. D. M. Ginsberg. Singapore: World Scientific, 1992, {\bf Vol.3}, 285-361.
\bibitem{TS} T. Timusk and B. Statt. Rep. Prog. Phys. {\bf 62}, 161 (1999).
\bibitem{P} A. A. Pashitskii, Low Temp. Phys. {\bf 21}, 763 (1995); and Low Temp. Phys. {\bf 21},  837 (1995).
\bibitem{Max} E. G. Maksimov, Physics-Uspekhi {\bf 43}, 965 (2000).
\bibitem{LL} B. Leridon, A. Defossez, J. Dumont, J. Lesueur, and. J. P. Contour,  Phys. Rev. Lett. {\bf 87}, 197007(1- 4) (2001).
\bibitem{St} E. Stajic, A. Iyengar, K. Levin, B. R. Boyce, and T. R. Lemberger, Phys. Rev. B {\bf 68}, 024520 (2003).
\bibitem{Pr} D. D. Prokof'ev, M. P. Volkov, and Yu. A. Bojkov, Fiz. Tverd. Tela, {\bf 45}, 11 (2003).
\bibitem{Wang} K. W. Wang, W. Y. Ching, Physica C, {\bf 416}, 47 (2004).
\bibitem{Za} R. O. Zaitsev, Zh. Eksp. Teor. Fiz. (JETF), 125, 891
(2004).
\bibitem{Ino} D.S. Inosov, J.T. Park, A. Charnukha, Yuan Li, A.V. Boris, B. Keimer, V. Hinkov,
Phys. Rev. B83, 214520 (2011).
\bibitem{Fish} Q. Fischer, M. Kugler, I. Maggio-Aprile, Christophe Berthod, Rev. Mod. Phys. 79,
353 (2007).
\bibitem{Im} M. Imada, A. Fujimori, Y. Tokura, Rev. Mod. Phys.- 1998. {\bf 70}, 1040 (1998).
\bibitem{Zu} E.E. Zubov, Physica C, {\bf 497}, 67 (2014).

\end{thebibliography}
\end{document}